\documentclass[12pt]{article}

\usepackage{newtxtext,newtxmath}

\usepackage{graphicx}
\usepackage[T1]{fontenc}
\usepackage[sort&compress,numbers,super]{natbib}
\usepackage{caption}
\usepackage[letterpaper,margin=1in]{geometry}

\renewenvironment{abstract}
	{\quotation}
	{\endquotation}

\date{}

\makeatletter
\renewcommand{\fnum@figure}{\textbf{Figure \thefigure}}
\renewcommand{\fnum@table}{\textbf{Table \thetable}}
\makeatother

\usepackage{url}
\usepackage{hyperref}

\def\scititle{
	Light-based electron aberration corrector
}
\title{\bfseries \boldmath \scititle}

\author{
	Marius Constantin~Chirita Mihaila$^{1\ast}$, Petr Koutenský$^{1}$, Kamila Moriová$^{1}$,\\ and
	Martin~Kozák$^{1\ast}$\and
	\small$^{1}$Department of Chemical Physics and Optics, Faculty of Mathematics and Physics, Charles University,\\ \small Ke Karlovu 3, Prague CZ-12116, Czech Republic\and
	\small$^\ast$Corresponding authors. Email: marius.chirita@matfyz.cuni.cz, m.kozak@matfyz.cuni.cz
}

\begin{document} 

\maketitle

\begin{abstract} \bfseries \boldmath
Achieving atomic resolution in electron microscopy has historically been hindered by spherical aberration, a fundamental limitation of conventional electron lenses. Its correction typically requires complex assemblies of electromagnetic multipoles. Here, we demonstrate that third-order spherical aberration in a cylindrically symmetric electron lens with an associated aberration coefficient of $C_\text{s}\approx 2.5\,\mathrm{m}$ can be compensated via interaction with a shaped light field. By analyzing distortions in high-magnification point-projection electron images of optical standing waves, we quantify the spherical aberration before and after light-induced correction. The spatial distribution of the correction optical field is precisely characterized in situ using ultrafast four-dimensional scanning transmission electron microscopy utilizing the transverse deflection of electrons induced by the optical ponderomotive force. Such a combined characterization and correction approach introduces a new paradigm for optical control in electron beam shaping and opens a pathway towards compact and tunable light-based aberration correctors for high-resolution electron microscopy.
\end{abstract}

\noindent
Spherical aberration in an optical system causes rays that pass farther from the optical axis to be focused closer to the lens than those passing near the axis, resulting in a blurred focal spot and limited resolution~\cite{Scherzer1936}. In contrast to light optics, where spherical aberration results from the curved shape of lens surfaces and chromatic aberration stems from wavelength-dependent variations in refractive index, both of which can be mitigated through tailored lens shapes or multi-element designs, the situation in electron optics is fundamentally different. Scherzer’s theorem predicts that electrostatic and magnetostatic fields in rotationally symmetric electron lenses cannot be configured to eliminate spherical aberration~\cite{Scherzer1936}.

For half a century, efforts to correct these aberrations have met limited success. Early multipole corrector designs demonstrated the possibility of compensation, but did not translate into actual improvements in resolution due to mechanical instability, electromagnetic interference, and imperfect alignment~\cite{rose2009historical}. The first successful implementation of spherical aberration correctors made in 1990~\cite{haider1998electron} revolutionized electron microscopy by enabling \textnormal{sub-\AA{}ngström} spatial resolution. Since then, direct imaging of materials with atomic resolution has advanced materials science, nanotechnology, and structural biology, and has improved precision in electron beam lithography for the semiconductor industry~\cite{urban2008studying}. However, the performance of electron aberration correctors is limited to the compensation of low-order aberrations because they do not offer arbitrary phase-shaping capabilities.

In light optics, the advent of programmable and adaptive devices, particularly spatial light modulators (SLMs), has enabled precise and dynamic control of optical wavefronts~\cite{maurer2011spatial}. In electron optics, however, phase shaping technologies have emerged only recently~\cite{grillo2014,shiloh2018spherical,VERBEECK201858,ribet2023design}, operating through an analogy to refractive-index modulation in light optics. Here, the electrons must propagate through a solid-state structure, leading to unavoidable scattering losses and degradation of the beam quality. To solve these limitations, one can utilize the interaction between electrons and photons to modulate the phase of electron beams either in the vicinity of a nanostructure~\cite{barwick2009photon,feist2015quantum,vanacore2018attosecond,vanacore2019ultrafast,konevcna2020electron,ben2021shaping,shiloh2021electron,henke2021integrated,dahan2021imprinting,feist2022cavity,madan2022ultrafast,tsesses2023tunable,garcia2023spatiotemporal,gaida2023lorentz,synanidis2024quantum,fang2024structured,ferrari2025realization} or in free space~\cite{freimund2001observation,freimund2002bragg,hebeisen2008grating,kozak2018inelastic,Kozak2018,schwartz2019laser,axelrod2020observation,tsarev2023nonlinear,lin2024ultrafast,velasco2025free}. Free-space modulation of electron wavefronts based on ponderomotive interaction offers distinct advantages, inherently minimizing decoherence and avoiding electron loss. Moreover, combining the light shaping capabilities of SLMs with the electron phase modulation has recently enabled adaptive phase control of electrons~\cite{de2021optical,chirita2022transverse}. 

The underlying mechanism of the free-space electron–light interaction applied in this work can be understood either semi-classically, as a phase modulation of an electron wave by the electromagnetic field of light, or quantum mechanically, as stimulated Compton scattering~\cite{de2021optical,chirita2022transverse}. In the former picture, the electron wave propagating through the classical electromagnetic fields acquires a phase modulation proportional to the local light intensity. In the latter picture, the electron coherently acquires a position-dependent phase through the simultaneous absorption and emission of photons~\cite{de2021optical}. One of its most compelling applications is the dynamic correction of aberrations in electron lenses, a topic that has attracted significant recent theoretical interest~\cite{de2021optical,uesugi2021electron,uesugi2022properties,ChiritaMihaila:25,mihaila2025light,nekula2025laser,guo2025spatiotemporal,uesugi2025crossed}.

While the proof-of-concept experiments~\cite{chirita2022transverse} demonstrated the transfer of spatial modulation from light to electrons, its application to electron aberration correction requires two major issues to be resolved.

\begin{figure}[]
   \centering      
   \includegraphics[width=16.0cm]{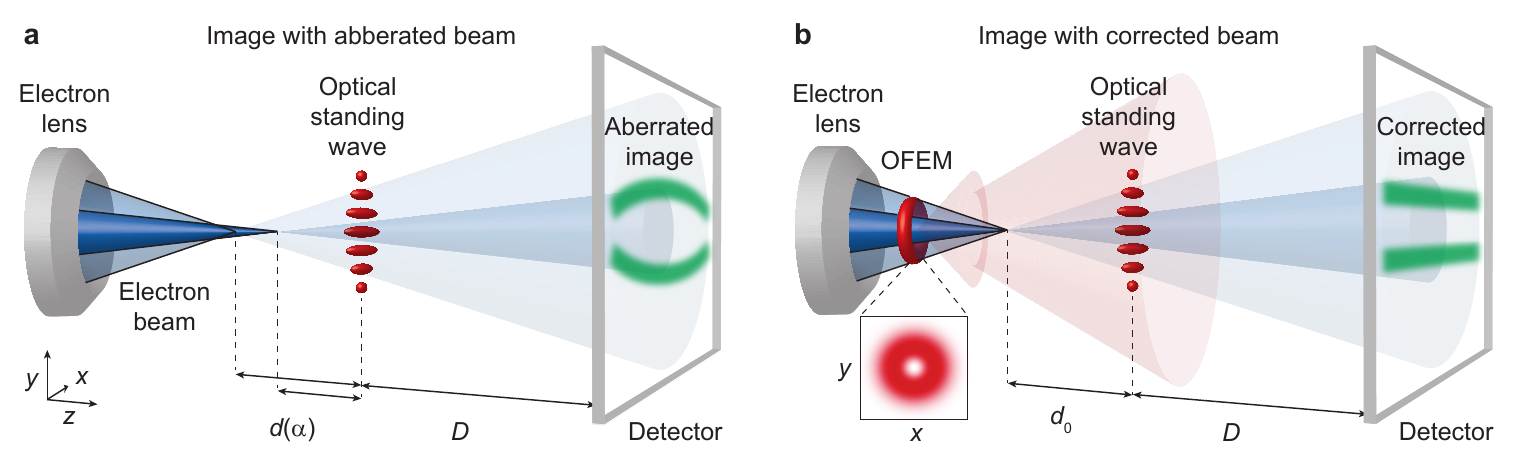} \caption{\textbf{Principle of light-based corrector of spherical aberration of a focused electron beam.}
            (a) The focal distance in an aberrated beam is a function of the convergence angle of rays. A point-projection image of equidistant fringes of an optical standing wave generated on the detector is distorted due to angle-dependent magnification $M(\alpha)=D/d(\alpha)$, where $\alpha$ is the deviated angle. 
            (b) Application of a Laguerre-Gaussian beam of charge one (its intensity distribution is shown in the inset) upstream the focus (OFEM) leads to correction of the spherical aberration. All rays are focused to a common focal point resulting in constant magnification across the detector and undistorted image of equidistant straight fringes. The distance $d_0$ between the geometrical focal point and the optical standing wave does not depend on the angle anymore. The electron beam and the laser fields are pulsed. The semi-transparent red cone in (b) schematically represents the converging Laguerre–Gaussian laser beam that forms the OFEM. }
   \label{Figure1}
\end{figure}
The first critical requirement for aberration correction demonstration is a precise method that enables one to determine the coefficient of spherical aberration $C_\text{s}$ during the corrector operation. Aberrations in electron lenses can be characterized using diffractogram analysis~\cite{saxton2000new}, Ronchigrams~\cite{cowley1979adjustment}, and, when these are not applicable, shadow imaging with fine gratings serves as a valuable alternative~\cite{rempfer1997simultaneous,ChiritaMihaila:25}. In our study, we employ a novel method, point-projection imaging of an optical standing wave, which acts as an etalon sample composed of light.  The distortion of the resulting image is a direct measure of $C_\text{s}$. At the same time, the interaction of electrons with the optical standing wave occurs in vacuum, preventing additional scattering or electron loss.  However, the presence of other types of aberration may complicate the analysis.

The second challenge is the precise in situ characterization of the spatial profile of the optical intensity that modulates the electron beam. Visualizing weak phase objects in electron microscopy relies on converting subtle phase shifts into measurable intensity variations. Techniques such as off-axis electron holography, Zernike phase contrast, in-line phase contrast, and laser-based phase modulation have been developed to achieve this transformation~\cite{axelrod2024modern}. These advanced methods are especially promising for imaging biological specimens, where enhanced phase sensitivity is crucial for revealing fine structural details without the need for staining or exposure to high electron doses. We introduce here the Ultrafast four-dimensional scanning transmission electron microscopy  (U4DSTEM) method, which enables direct in situ mapping of light-induced electron phase modulation at the interaction plane with nanometer-scale resolution provided by the focused electron beam, and allows reconstruction of the underlying optical intensity profile.

\section*{Results}
\addcontentsline{toc}{section}{Configuration for light-based aberration correction}

We experimentally demonstrate the correction of spherical aberration of an electron beam induced by focusing system of a scanning electron microscope column equipped with a field-emission source. The aberration and its correction are shown via monitoring a highly magnified point-projection image of an optical standing wave, generated by two counter-propagating laser pulses in the perpendicular direction with respect to the electron beam, which serves as an etalon sample. The aberration is corrected via spatial phase modulation of the electron beam induced by interaction
with a shaped pulsed light beam, propagating in the opposite direction to the electrons, which we refer to as the optical field electron modulator (OFEM). Specifically, we apply Laguerre–Gauss (LG) mode of charge one. Although the LG beam acts similarly to a convex lens close to the center of the beam, it fundamentally differs from conventional electron lenses by introducing strong spherical aberration of negative sign enabling the compensation of spherical aberration induced by the traditional electron optics.

Figure~\ref{Figure1} illustrates the experimental setup of a light-based electron aberration corrector (a detailed description of individual parts of the setup can be found in the Methods section and Extended Data Fig. 1). Figure~\ref{Figure1}a shows the point-projection microscopy of the optical standing wave measured with an aberrated electron beam. The standing wave results from an interference of two counter-propagating pulsed Gaussian beams,  which can be approximated as plane waves near the focus. This results in a pattern resembling a perfect etalon sample, with straight, parallel optical fringes spaced by half the wavelength of the light used to generate the standing wave. When the electrons propagate through the optical standing wave, the rays become focused in one direction in the local minima of the ponderomotive potential and defocused in its maxima. The optical standing wave thus acts as an equidistant series of cylindrical lenses for electrons. As a result, when the intensity of the standing wave is appropriately chosen to focus the electrons in the detector plane, fringes appear at the downstream detector. We note that the optical standing wave can be viewed as a one-dimensional lens array which, together with the 2D detector, offers an analogy to an electron-optical counterpart of the Shack–Hartmann wavefront sensor used in laser optics~\cite{platt2001history}. Spherical aberration with positive sign, which is present in all electromagnetic lenses with cylindrical symmetry~\cite{rose2009historical}, causes a shift of the focal point closer to the electron lens with increasing angle between the electron trajectory and the beam axis. This leads to angle-dependent magnification $M(\alpha)=D/d(\alpha)= D/(d_0 + C_\mathrm{s}\alpha^2)$, where $D$ is the distance between the optical standing wave and the detector, $d_0$ is the distance between the geometrical focal spot and the optical standing wave, and $d(\alpha)$ represents the angle-dependent path length between the focal spot and the standing wave, incorporating spherical aberration. The image acquired with the aberrated electron beam thus shows curved fringes with their curvature directly related to the value of the spherical aberration coefficient $C_\text{s}$. Once the aberration correction by OFEM is applied (see Figure~\ref{Figure1}B), all electron rays are focused to a single focal point leading to straight fringes at the detector.

\begin{figure}[t]
   \centering      
   \includegraphics[width=16.0cm]{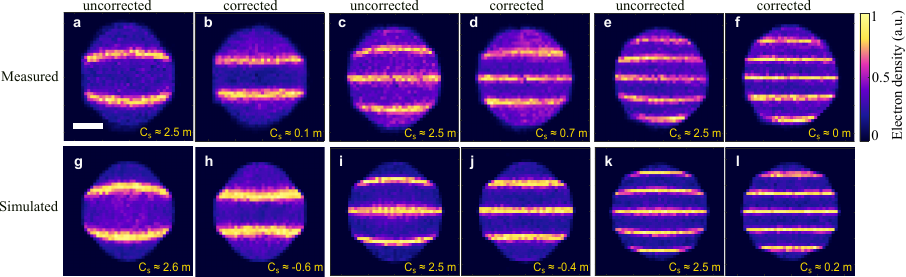} \caption{\textbf{Electron point projection images of optical standing wave acquired with electron beam with and without spherical aberration.} 
   Upper row: Experimental images measured with the aberrated electron beam (panels a,c,e) are compared with the images acquired with the electron beam corrected by OFEM (b,d,f) (scale bar: $255\,\mu\mathrm{m}$). The corresponding values of the spherical aberration coefficient obtained by the procedure described in Supplementary Materials are shown in the corresponding panels. The data were obtained with three slightly different focal distances of the objective lens leading to different magnifications. Lower row: Numerical simulations of images acquired with aberrated (panels g,i,k) and corrected electron beam (panels h,j,l) calculated with the same parameters, which were used in the experiments.}
   \label{fig:Aberration corrected images}
\end{figure}

In Fig.~\ref{fig:Aberration corrected images}, we present the experimental results (panels a–f) alongside the corresponding simulations (panels g–l). The images were acquired using an aberrated electron beam (panels a,c,e,g,i,k) and an aberration-corrected beam (panels b,d,f,h,j,l). Data was collected at three different magnifications, approximately 1288x, 937x, and 714x, achieved by varying the working distance of the microscope (distance $d_0$).

The fringes were fitted with an analytical function to obtain the value of $C_{\mathrm{s}}$, which is $C_{\mathrm{s}} \approx 2.5\,\pm 0.1\,\mathrm{m}$ for the aberrated beam and $C_{\mathrm{s}} \approx 0.1 \pm 0.1\,\mathrm{m}
$ for the corrected beam at highest magnification. The uncertainty in $C_\mathrm{s}$ is estimated based on the standard error of the fitted fringe positions, calculated from the residuals of the fit and the sensitivity of the model to changes in $C_\mathrm{s}$ (see Supplementary Materials for details). We note that for each image, the magnification, achieved by shifting the focal spot of the electron beam, also changes the beam size at the interaction plane with the OFEM. As a result, both the defocus of the laser beam and its temporal overlap with the electron pulse must be carefully readjusted. The residual $C_\mathrm{s}$ observed in the experimental data of Fig.~\ref{fig:Aberration corrected images}d likely reflects imperfect laser alignment under these conditions. The results confirm that the OFEM with LG beam effectively compensates the spherical aberration of the electron beam. 

One of the crucial requirements for successful implementation of light-based aberration correctors in electron microscopy is a precise characterization of the spatial distribution of phase modulation induced to the electrons. To obtain this information directly in situ, we turned off the optical standing wave, focused the electron beam on the OFEM plane, and used U4DSTEM imaging~\cite{koutensky2025ultrafast}. We map the transverse momentum transfer imparted by the OFEM by scanning the electron focal spot and measuring the center of mass of electron distribution on the detector. Because the transverse momentum transfer is directly proportional to the gradient of the optical intensity integrated along the electron trajectory, such measurement represents a direct spatial mapping of the strength of the interaction between electrons and light.

\begin{figure}[]
   \centering      
   \includegraphics[width=12cm]{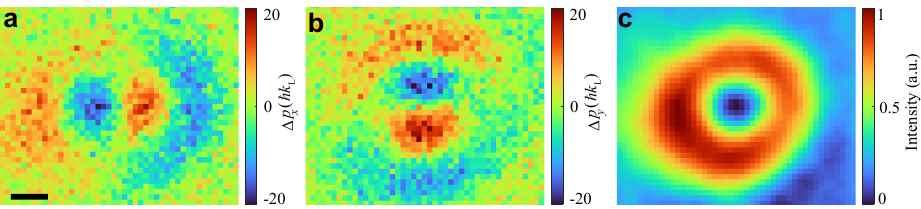} \caption{\textbf{Spatial distribution of the electron phase modulation induced by the light-based aberration corrector.} Panels a and b show the measured $x$ and $y$ components of the transverse momentum change of the electrons as a fuction of the electron beam position in the OFEM plane, respectively, obtained by U4DSTEM technique (scale bar $2\,\mu\mathrm{m}$). Panel c shows the reconstructed profile of the ponderomotive potential integrated along the electron trajectory of the LG mode obtained from the data in a,b.}
   \label{fig:Donut Gradient}
\end{figure}
From 2D gradient maps of the $x$- and $y$-components of the transverse momentum change of the electrons shown in Figures~\ref{fig:Donut Gradient}a,b we reconstructed the spatial profile of the phase modulation imprinted to the electron beam by OFEM (Fig.~\ref{fig:Donut Gradient}c). The reconstruction was performed by applying a Fourier-based Poisson solver to the measured beam shift gradients~\cite{bhat2008fourier} (see supplementary materials). The waist of the LG beam, \( w_{\mathrm{LG}} = 4.45\,\mu\mathrm{m} \), was extracted by fitting the reconstructed intensity distribution shown in Figure~\ref{fig:Donut Gradient}c, and subsequently used in the calculation of the ponderomotive phase shift in Eq.~\ref{eq:phase_shift} for all simulated images. These results demonstrate the ability of U4DSTEM to resolve light-induced momentum distributions and reconstruct optical field profiles with spatial resolution, which is not limited by light diffraction as in the case of optical imaging systems~\cite{chirita2022transverse}, but only by the resolution of the electron microscope itself. We note  that the resolution in Figure~\ref{fig:Donut Gradient}c is limited by the pixel size of 300 nm, but it can go down to $\approx 20\,\mathrm{nm}$~\cite{koutensky2025ultrafast}. The deviation from ideal LG beam observed in panel Fig.\ref{fig:Donut Gradient}c arises from aberrations induced in the optical setup (mirror with a hole, focusing lens). However, the inner part of the intensity distribution with a radius $< 3\,\mu\mathrm{m}$ that interacts with the electrons remains almost undistorted.
\begin{figure}[h]
   \centering      
   \includegraphics[width=7.0cm]{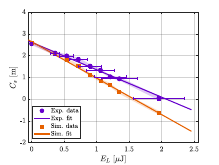} \caption{ \textbf{Electron spherical aberration coefficient as a function of OFEM pulse energy.} Panel shows the $C_{\mathrm{s}}$ coefficients determined from the experimental images (violet circles) and simulated images (orange squares) as a function of OFEM pulse energy. The shadow regions represent the accuracy of $C_s$ determination which are estimated to $\sim$0.1 m (see supplementary materials for details). The purple error bars indicate the estimated uncertainty in determining the laser pulse energy at the OFEM plane.}
   \label{Figure3}
\end{figure}

To quantify the OFEM tunability, we varied the pulse energy $E_\text{L}$ used for compensation. In Figure ~\ref{Figure3} we show the extracted $C_\text{s}$ values from the experimental and simulated images with the highest magnification. The full $C_{\mathrm{s}}$ correction is reached with a pulse energy of $E_L \approx 2\,\mu\mathrm{J}$. Details on the $C_{\mathrm{s}}$ extraction procedure and the experimental images used to obtain data in Fig.~\ref{Figure3} are provided in Supplementary Materials and figure~S2. The residual $C_\mathrm{s}$ observed in the simulations after correction is primarily attributed to an estimated 20$\%$ uncertainty in the laser pulse energy delivered to the OFEM plane, which also accounts for the difference in slopes between theory and experiment in Fig.~\ref{Figure3}, as both effects scale with the same factor. Furthermore, at lower image magnifications, the reduced pixel resolution limits the precision of curvature fitting in point-projection electron images, contributing to the observed undercompensation.

\section*{Numerical modeling}
\addcontentsline{toc}{section}{Numerical modeling}

To quantitatively interpret the experimental results, we use a ray-optics model that captures the propagation of the aberrated electron beam and its interaction with both the OFEM and the optical standing wave. In this approach, the electron beam is treated as an ensemble of classical rays, and wave interference or coherence effects are neglected. The ray tracing algorithm is initialized at the OFEM plane, with the initial propagation angles of the electrons given by $\theta_0(x) = -x / d_c$ and $\theta_0(y) = -y / d_c$, where $d_c$ denotes the distance from the OFEM plane to the geometrical focus of the electron beam, and $x$ and $y$ are the transverse spatial coordinates within the circular beam profile. The propagation of rays between the interaction planes and the detector is modeled using the free-space propagation matrix formalism. The local change in propagation direction is calculated from the phase gradient using the relation \(\delta\theta_{\mathrm{i}} = \frac{1}{k_e} \nabla\varphi_{\mathrm{i}}\), where \(\mathrm{i}\) denotes the contribution from spherical aberrations, the OFEM, or the optical standing wave.

Here, $k_e=2\pi/\lambda_e$ is the electron wavenumber, with $\lambda_e = h/(\gamma m_e v)$ denoting the relativistic electron wavelength. In this expression, $v$ is the electron velocity, $m_e$ is the electron mass, and $\gamma = 1/\sqrt{1-v^2/c^2}$ is Lorentz factor accounting for relativistic effects.

The wavefront error of the electron beam due to the spherical aberration is defined as~\textcolor{blue}{\cite{Scherzer1936}}:
\begin{equation}
    \varphi_{\mathrm{a}} (\theta(x,y))= -\frac{\pi}{2\lambda_e}C_{\mathrm{s}}\theta^4(x,y),
    \label{eq: Aberration function wave}
\end{equation}
where $\theta(x,y)$ is the semi-convergence angle and $C_{\mathrm{s}}$ is the spherical aberration coefficient.

The analytical expression for the phase shift acquired by an electron interacting with a counter-propagating LG laser focal spot is given by~\cite{de2021optical,chirita2022transverse}:
\begin{equation}
\begin{aligned}
    \varphi_{\mathrm{l}}(x,y)
    &\approx -\frac{\alpha}{2\pi (1+ \beta)} \frac{E_L \lambda_L^2}{E_e} 
    \frac{ g_{\mathrm{LG}}^2(x,y)}{\int_{-\infty}^{\infty} \int_{-\infty}^{\infty} \mathrm{d}x\, \mathrm{d}y\, g_{\mathrm{LG}}^2(x,y)}.
\end{aligned}
\label{eq:phase_shift}
\end{equation}

The parameters in the expression are defined as follows: $\alpha$ is the fine-structure constant, $\beta = v/c$ represents the electron’s velocity normalized to the speed of light, and $\lambda_L$ denotes the laser wavelength. The relativistic energy of the electron is given by $E_e = \gamma m_e c^2$. Furthermore, the spatial distribution of the laser intensity is given by $g_{\mathrm{LG}}^2(x, y) = (\rho^2 / w_{\mathrm{LG}}^2) \exp(-2\rho^2 / w_{\mathrm{LG}}^2)$, where $\rho = \sqrt{x^2 + y^2}$ denotes the radial coordinate.

When the electrons propagate through the optical standing wave formed by two counterpropagating Gaussian laser pulses propagating in $x$-direction, they acquire a phase shift that can be calculated as~\cite{hebeisen2008grating}:
\begin{multline}
     \varphi_{g}(x', y'=0, z'=0, t) = \frac{e^2 \lambda_L^2 I_0}{16 \pi^2 m_e \epsilon_0 c^3 \hbar} \int_{-\infty}^{\infty}  
     \left[ \exp\left(-\frac{\left(t - \frac{x'}{c}\right)^2}{2 w_t^2}\right) + \exp\left(-\frac{\left(t + \frac{x'}{c}\right)^2}{2 w_t^2}\right) \right. \\
    \left. + 2 \exp\left(-\frac{t^2}{2 w_t^2}\right) \exp\left(-\frac{x'^2}{2 w_t^2 c^2}\right) \cos(2k_Lx') \right] \, dt.
    \label{phi grating}
\end{multline}

Here, $I_0$ denotes the peak intensity of each individual laser pulse, $k_L = 2\pi/\lambda_L$ is the laser wavenumber, and the pulse duration is given by $\tau_l =2\sqrt{\mathrm{2ln2}}w_t$ and $x'$, $y'$ and $z'$ are the spatial coordinates in the plane of the optical standing wave, which we refer to as the sample plane (SP). The first two terms in the integral correspond to the individual counterpropagating pulses giving only a spatially constant phase offset, while the third term represents the resulting optical standing wave.

\section*{Discussion}
\addcontentsline{toc}{section}{Concluding remarks}

In this work, we demonstrate optical correction of 3rd-order spherical aberration in a conventional electron lens using a Laguerre–Gaussian (LG) beam with charge one. The method can be extended to correct not only 3rd-order spherical but also higher-order electron aberrations by employing a gradient descent optimization algorithm to iteratively refine the laser wavefront using an SLM capable of both phase and amplitude modulation. The optimization process can be guided by the acquired electron image itself, enabling real-time autotuning of the optical field to maximize image quality~\cite{dellby2001progress}.

The efficiency of spherical aberration correction depends strongly on the spatial extent of the shaping region. Specifically, the laser pulse energy required to compensate a certain value of $C_\mathrm{s}$ scales quadratically with the electron beam radius in the interaction plane $w_e$, which also determines the radius of the laser beam used for OFEM, $w_{\mathrm{LG}}$ (the ratio $w_e / w_{\mathrm{LG}} \approx 0.67$ should be maintained for compensation of the third-order spherical aberration without introducing higher order aberration terms). This is a consequence of the linear dependence of the induced ponderomotive phase shift on the local intensity of light.

The transverse coherence of the electron beam is not altered by the interaction because it is coherent and elastic. For integration into a transmission electron microscope, the OFEM can be positioned upstream of the sample, as experimentally demonstrated in previous studies that achieved pre-sample electron modulation using optical fields near thin films~\cite{fang2024structured,ferrari2025realization}, enabling the aberration correction without directly perturbing the specimen.

Standard high-resolution scanning electron microscopes achieve a best spatial resolution of about 0.5\,nm, which is insufficient to resolve individual atoms in solid-state samples. Assuming a maximum electron energy of 30\,keV, the corresponding de~Broglie wavelength is approximately 7\,pm. With an electron beam divergence angle of 50\,mrad, the theoretical resolution limit is about 70\,pm, which has previously only been achieved using a multipole aberration corrector\cite{sawada2015atomic}.
As ultrafast electron microscopes~\cite{zewail2010four,morimoto2018diffraction} combine atomic spatial resolution with femtosecond temporal precision to visualize dynamic processes at the space-time scale, they could greatly benefit from this light-based aberration correction approach, which offers a compact and tunable alternative to traditional multipole correctors.

The crucial points for applicability of the light-based electron optical correctors in standard electron microscopes are the spatial stability of the corrector and the possibility to work in the continuous regime. The spatial stability of OFEM is determined by the combination of the pointing stability of the optical beam and the mechanical stability of the focusing lens and other optical components. For reaching sub-nanometer spatial resolution, the OFEM also has to be spatially stabilized with the same precision, which represents a major challenge for future research. Continuous wave (CW) operation requires enhancing the electron–light interaction strength, which can be achieved using a resonant optical cavity to accumulate sufficient optical intensity~\cite{schwartz2019laser}. Although cavity powers on the order of 100 kW have been previously demonstrated, the power required to correct spherical aberrations in transmission electron microscopes could be reduced by aligning the cavity coaxially with the electron beam. In the current configuration, the effective interaction distance is $d_{\mathrm{int}} \approx 26\,\mu\mathrm{m}$, is limited by the pulsed nature of the electrons and light~\cite{chirita2022transverse}. In contrast, CW operation would enable an extended interaction distance. Another parameter influencing the OFEM efficiency via the amplitude of the phase shift induced by the ponderomotive interaction is the wavelength of light used in OFEM. Longer wavelengths enhance the interaction strength for a fixed beam waist, although maintaining a constant beam radius with increasing wavelength requires an increase of numerical aperture of the focusing setup, which may pose practical challenges.  We note that the selection of wavelength and numerical aperture also influences the spatial resolution of OFEM, which is ultimately limited by light diffraction.

In summary, we have demonstrated that the dominant electron lens aberration, specifically the third-order spherical aberration, can be corrected via free-space interaction with a shaped light beam. 

In addition, we introduced the U4DSTEM technique that allows in situ characterization of the phase profile imprinted on the electrons, which is essential for further development of any structured light-based electron phase modulators. Furthermore, the optical standing wave enables in situ calibration and assessment of electron optical distortions using a known optical phase etalon, providing a powerful tool for precise wavefront metrology within these setups. 

Unlike conventional aberration correctors requiring multiple lenses, which also offer real-time feedback-based aberration compensation, our single-plane light-based approach may offer a pathway toward miniaturized aberration correction systems. With continued progress in adaptive optics and AI-driven feedback control, dynamically tunable, light-based optical elements may become integral to next-generation electron microscopy systems, enabling real-time, reconfigurable beam shaping with unprecedented precision and flexibility.

\clearpage 

%
\bibliography{literatur} 
\bibliographystyle{naturemag}

%
%
%
%
%
%

\newpage

\subsection*{Methods}

In our experimental setup, illustrated in Extended Data Fig. 1, we generate ultrashort electron pulses via laser-triggered photoemission from a Schottky field emission gun integrated into a Verios 5 UC scanning electron microscope (Thermo Fisher Scientific). We used the second harmonic $(515\,\mathrm{nm})$ of a femtosecond ($400\,\mathrm{fs}$, $50\,\mathrm{kHz}$) laser pulse to trigger photoemission from a Schottky field emitter, producing single electron pulses that reach the SP with a duration of approximately $680\,\mathrm{fs}$ at an energy of $20\,\mathrm{keV}$ with a maximum semi-convergence angle of $\alpha_{\mathrm{max}} = 5\,\mathrm{mrad}$$^{57}$. The transverse spatial properties of the pulsed electron beam do not change significantly compared to the continuous emission mode.

As electrons propagate towards the sample plane (SP), they accumulate spherical aberration, with the probe-forming lens located approximately~$5\,\mathrm{cm}$ upstream of the SP. Due to the absence of an aperture, the electron beam radius at the lens is estimated to be~$225\,\mu\mathrm{m}$, which contributes to the strength of the 3rd order spherical aberration. The resulting wavefront distortion is characterized by a spherical aberration coefficient of $C_{\mathrm{s}} \approx 2.5\,\mathrm{m}$. To compensate for aberrations, the electron wavefront is phase-modulated at the OFEM plane through interaction with a LG beam, generated by a spatial light modulator (SLM; Holoeye Photonics, Pluto 2.1). The converging electron beam, with a radius of approximately~$3\,\mu\mathrm{m}$, illuminates only the central region of the LG mode. After this initial interaction, the electrons pass through a crossover and propagate to the SP, where they interact with a second optical field in the form of a standing wave, which is used as an etalon sample. The standing wave is generated by focusing two counter-propagating pulsed laser beams with the wavelength of 1030 nm and pulse duration of 400 fs. The beam radius in the focus is $w_0 = 10\,\mu\mathrm{m}$. The resulting electron image of the standing wave is recorded on a position-sensitive detector (Timepix 3, Advascope) located~$17\,\mathrm{cm}$ downstream of the SP. The distance between OFEM and SP is approximately~$0.74\,\mathrm{mm}$, and all experimental point-projection images were acquired with an exposure time of 5~seconds.

The standing wave serves as an etalon sample with exact spacing of its interference maxima and perfectly flat phase-front (we only observe a small part of the interference pattern of the optical standing wave with the size of approximatively $700\,\mathrm{nm}$), enabling imaging near the electron cross-over at high magnifications, where spherical aberrations are most pronounced, without interfering with the OFEM. The $\lambda_L/2$ periodicity of the standing wave creates high spatial phase gradients, resulting in pronounced fringes at the detector plane using $E_L  < {1}{\mu J}$ per counterpropagating laser pulse and a numerical aperture NA$= 0.08$.

The OFEM is brought on-axis with the electron beam using a pellicle beam splitter (BP145B3, Thorlabs, not shown in Extended Data Fig. 1) with a hole in the center. This hole allows the electron beam to pass through unobstructed, while enabling co-linear alignment of the OFEM. The hole is fabricated in-house by diffracting the laser focus (NA$={0.13}{}$) into a circular pattern with a radius of $150\,\mu\mathrm{m}$ using the spatial light modulator (SLM) and applying pulses with an energy of $6\,\mu\mathrm{J}$. The NA of the OFEM is ${0.13}{}$.

The timing between the OFEM and the electron pulse is adjusted using the first delay stage. The second delay stage controls the relative timing of the counterpropagating laser pulses that form the standing wave, whereas both the second and third delay stages are used to synchronize the standing wave with the arrival of the electron pulse at the SP.

\subsection*{Data availability}
The data that support the findings of this paper are available via Zenodo at
\url{https://doi.org/10.5281/zenodo.16532408} (ref.~$^{58}$).

\subsection*{Code availability}
All computational codes used to generate the data presented in this study are available from the corresponding author upon reasonable request.

\subsection*{References}

\noindent
\hangindent=1.8em
\hangafter=1
[57] Moriová, K., Koutenský, P., Chirita-Mihaila, M.-C. \& Kozák, M. Temporal characterization of femtosecond electron pulses inside ultrafast scanning electron microscope. \textit{Review of Scientific Instruments} \textbf{96} (2025).

\noindent
\hangindent=1.8em
\hangafter=1
[58] Chirita Mihaila, M. C., Koutenský, P., Moriová, K. \& Kozák, M.
Data for ‘Light-based electron aberration corrector’. Zenodo
\url{https://doi.org/10.5281/zenodo.16532408} (2025).

\subsection*{Acknowledgments}

We acknowledge fruitful discussions with A. I. C. Mihaila regarding the simulation work. We also acknowledge funding from the Czech Science Foundation (project number 22-13001K), Charles University (SVV-2023-260720, PRIMUS/19/SCI/05 and GAUK 90424) and the European Union (ERC, eWaveShaper, 101039339). This work was supported by the Ministry of Education of the Czech Republic, project TERAFIT (CZ.02.01.01/00/22\_008/0004594). M.C.C.M. acknowledges the support of the project MSCA Fellowships CZ – UK4 (CZ.02.01.01/00/22\_010/0013392) under the Programme Johannes Amos Comenius.

\subsection*{Author contributions}
MCCM and MK conceived the experiment. MCCM, MK, PK, and KM performed the experiments and analyzed the data. MCCM performed the calculations and simulations. MCCM and MK wrote the manuscript.
\subsection*{Competing interests}
There are no competing interests to declare.

\newpage

\begin{figure}[h]
   \centering      
   \includegraphics[width=8.0cm]{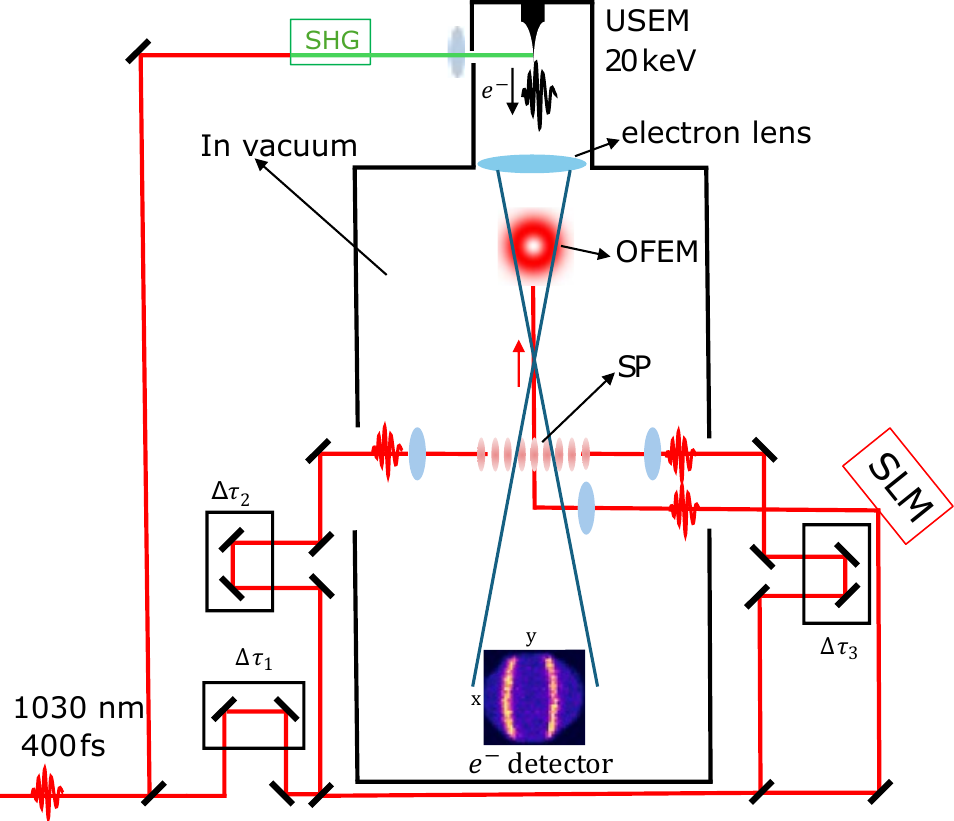} \caption*{\textbf{Extended Data Fig. 1|Schematic of the experimental setup for spherical aberration correction of electrons using light.} 20 keV pulsed electron beam ($\tau_e = 680\,\mathrm{fs}$) generated by linear photoemission process by the second harmonic frequency of a femtosecond laser ($\lambda_L = 515\,\mathrm{nm}$, $\tau_l = 400\,\mathrm{fs}$
) is focused by the objective lens of an electron microscope column. The aberrated electron beam interacts with a LG beam at the OFEM plane generated using a spatial light modulator (SLM). Subsequently, the electron pulses reach the sample plane (SP), where they interact with an optical standing wave, which introduces additional phase modulation. Point-projection electron images of the optical standing wave with high magnification (see inset) measured using a 2D hybrid-pixel detector are used to evaluate the spherical aberration coefficient of the electron beam, which corresponds to the curvature of the measured fringes at the detector. An effective correction at the OFEM plane results in straight fringes in the detected image, indicating that the magnification is uniform across all propagation angles. The schematic is not to scale.}
   \label{fig:Experimental Setup}
\end{figure}


\newpage


\renewcommand{\thefigure}{S\arabic{figure}}
\renewcommand{\thetable}{S\arabic{table}}
\renewcommand{\theequation}{S\arabic{equation}}
\renewcommand{\thepage}{S\arabic{page}}
\setcounter{figure}{0}
\setcounter{table}{0}
\setcounter{equation}{0}
\setcounter{page}{1} 


\begin{center}
\section*{Supplementary Materials for\\ \scititle}

Marius Constantin Chirita Mihaila$^{\ast}$, Petr Koutenský, Kamila Moriová, Martin Kozák$^{\ast}$
\small$^\ast$Corresponding authors. Email: marius.chirita@matfyz.cuni.cz, m.kozak@matfyz.cuni.cz\\
\end{center}

\subsubsection*{This PDF file includes:}
Supplementary Text\\
Figures S1 and S2\\

\newpage


\subsection*{Supplementary Text}

\subsubsection*{\label{sec:B} Calculating the spherical aberration coefficient of the electron beam}

\begin{figure*}[h]
   \centering      
   \includegraphics[width=16.0cm]{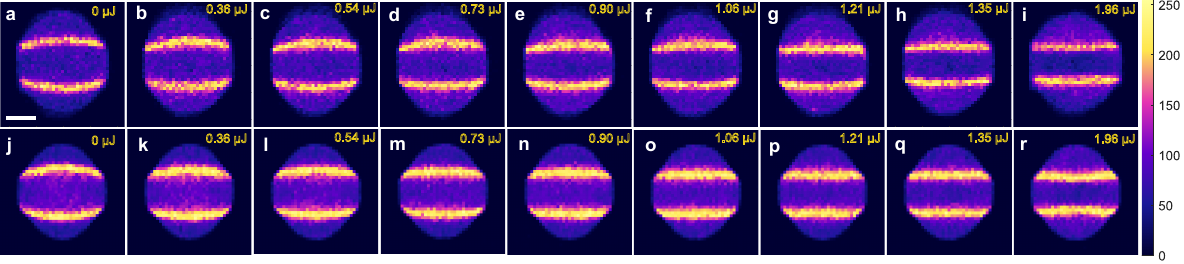} \caption{\textbf{Measurement of spherical aberration correction as a function of OFEM pulse energy.} Panels (a–i) show experimental electron point-projection images (scalebar ${510}{\mu m}$) of an optical standing wave acquired at increasing OFEM pulse energies. Panels (j–r) show the corresponding simulations. As the pulse energy increases, the curvature of the fringes progressively straightens, indicating effective aberration compensation. All images of the standing wave are shown in their raw form, without any image processing.
}
   \label{fig:power_dependance}
\end{figure*}

Figure~\ref{fig:power_dependance} presents a direct comparison between the experimental point-projection electron images (panels a–i) and the corresponding simulations (panels j–r) across a range of pulse energies. At low energies, the fringes appear notably curved due to the uncorrected spherical aberration. As the OFEM pulse energy increases, the curvature of these fringes progressively straightens, indicating that the aberration is being compensated. At a specific pulse energy, the fringes become nearly flat (panels I/R), marking the point of optimal aberration compensation. 

The pulse energies in the focal spot of the full LG beam corresponding to images a–i and j–r are, in units of ${}{\mu J}$: 0, 0.36, 0.54, 0.72, 0.90, 1.06, 1.21, 1.35, and 1.96. It should be noted that the electron beam interacts with only a fraction of the total pulse energy, as it is confined to the central region of the LG beam. 

The $C_{\mathrm{s}}$ was determined using the following procedure: The images were converted to grayscale and smoothed column-wise to enhance peak detection. Subpixel stripe positions were determined by fitting a 1D Gaussian to a 5-pixel window centered around the brightest peaks in each image column, enabling accurate localization of the top and bottom electron fringes. The transverse fringe positions were converted from pixel to metric units using the known pixel size of the detector. Quadratic functions were fitted to the top and bottom fringe positions as a function of horizontal coordinate, with the beam center, identified as the location of maximum fringe separation, establishing the origin of the coordinate system used in subsequent analysis.

\begin{figure*}[h]
   \centering      
   \includegraphics[width=16.0cm]{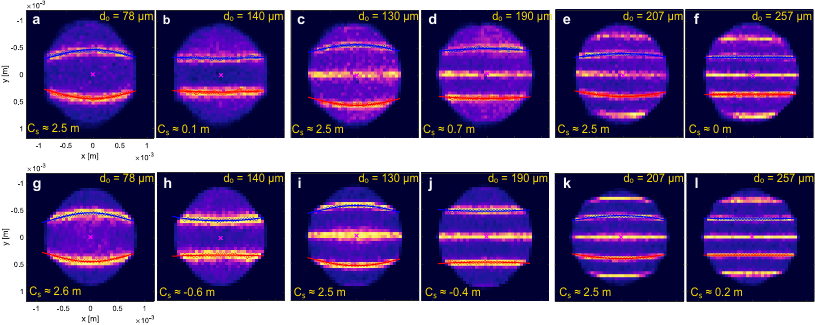} \caption{ \textbf{Electron fringe fit curvature before and after OFEM correction, shown for experimentally acquired images at different magnifications.}  Panels a and b show the experimental fringe patterns with the OFEM off and on, respectively, while Panels c and d present their corresponding simulations. In each case, the positions of the top and bottom fringes were fitted with cubic polynomials (Eq.~\ref{Eq:qubic}) to quantify the curvature and extract the $C_{\mathrm{s}}$ value. The slight asymmetry between the top and bottom curves in each image can be attributed to a small lateral displacement of the standing wave relative to the center of the electron beam. Note that for panels c to f, and i to l, the denominator of Eq.~\ref{Eq: Magnification} changes from $\lambda/2$ to $\lambda$.}
   \label{fig:Cs}
\end{figure*}

After centering, a second round of cubic fits (Eq.~\ref{Eq:qubic}) was performed to extract the curvature of the top and bottom fringes relative to the beam center; for each image, $C_{\mathrm{s}}$ was computed separately from both fringes and their mean value was recorded (see Fig.~\ref{fig:Cs}). This analysis was performed on experimental images a to i and simulated images j to r from Fig.~\ref{fig:power_dependance}, and the resulting $C_{\mathrm{s}}$ values are plotted in Fig.~\ref{Figure3} in the main manuscript.  To estimate the uncertainty of the $C_{\mathrm{s}}$, we fit the cubic function (Eq.~\ref{Eq:qubic}) independently to the top and bottom electron fringe positions. For each fit, the residuals are used to compute the mean squared error (MSE). The sensitivity of the model to $C_{\mathrm{s}}$ is evaluated numerically using finite differences. The standard error is then calculated as:
\begin{equation}
\sigma_{C_{\mathrm{s}}} = \sqrt{\frac{\mathrm{MSE}}{\sum_i \left( \frac{dy_i}{dC_s} \right)^2}
},
\end{equation}
where \( y_i \) (equivalent to \( y_{\text{det}} \) used later) is the model-predicted stripe position for the \( i \)-th data point, and \( \frac{dy_i}{dC_s} \) is its numerical derivative with respect to \( C_s \), computed via:
\begin{equation}
\frac{dy_i}{dC_s} \approx \frac{f(C_s + \delta, x_i) - f(C_s, x_i)}{\delta}.
\end{equation}
Here, \( f(C_s, x_i) \) is the cubic model evaluated at position \( x_i \), and \( \delta \) is a small perturbation.

The final uncertainty is obtained by combining the top and bottom errors via root-mean-square averaging:

\begin{equation}
\sigma_\mathrm{final} = \sqrt{\frac{\sigma_{C_{\mathrm{s}},\mathrm{top}}^2 + \sigma_{C_{\mathrm{s}},\mathrm{bottom}}^2}{2}}.
\end{equation}

The cubic fit function (Eq.~\ref{Eq:qubic}) is derived based on the following considerations. The radial distance from the center of the electron beam on the detector is defined as:
\begin{equation}
  r_{\text{det}} = \sqrt{x_{\text{det}}^2 + y_{\text{det}}^2},
\label{Eq:S4}
\end{equation}
where \(x_{\text{det}}\) and \(y_{\text{det}}\) are detector-plane coordinates of the fringes with respect to the center. Under the small-angle approximation, the angular deviation from the electron beam axis is
$\alpha = r_{\text{det}}/D$
with \(D\) denoting the distance from the optical standing wave to the detector.

The longitudinal shift of the focal distance due to spherical aberration can be expressed as:
\begin{equation}
\Delta z (\alpha) = C_{\mathrm{s}} \alpha^2.
\label{Eq:S5}
\end{equation}
With \(d_0\) as the distance from the standing wave to the geometrical electron cross over (focal point for on-axis rays), the magnification of two fringes displaced from the center becomes:
\begin{equation}
  M(\alpha) = \frac{D+d_0 + C_{\mathrm{s}} \alpha^2}{d_0 + C_{\mathrm{s}} \alpha^2} \doteq \frac{D}{d_0 + C_{\mathrm{s}} \alpha^2} = \frac{2y_{\text{det}}}{(\lambda_L/2)}.
  \label{Eq: Magnification}
\end{equation}

Here we used the fact that the distance between the geometrical cross over and the optical standing wave $d_0$ is much smaller than the distance between the optical standing wave and the detector $D$. The right hand side of Eq.~\ref{Eq: Magnification} corresponds to the experimentally measured magnification obtained by dividing the measured fringe separation ${2y_{\text{det}}}$ by the known period of the optical standing wave ${(\lambda_L/2)}$. The fit function used to evaluate the curvature of the fringes and retrieve the $C_{\mathrm{s}}$ coefficient is obtained by substituting Eqs.~\ref{Eq:S4} and~\ref{Eq:S5} into Eq.~\ref{Eq: Magnification}. The resulting equation has the form:
\begin{equation}
  2 C_{\mathrm{s}} y_{\text{det}}^3 + \left( 2 d_0 D^2 + 2 C_{\mathrm{s}} x_{\text{det}}^2 \right) y_{\text{det}} - \frac{\lambda_L}{2} D^3= 0.
  \label{Eq:qubic}
\end{equation}
The solution of cubic Eq.~\ref{Eq:qubic} is used to fit the experimentally measured positions of fringes in the detector plane. The electron defocus component associated with the LG beam results in a shift of the geometrical focal spot by only 60 $\mu$m upstream towards the objective lens. 

Furthermore, we did not observe a measurable contribution from fifth-order spherical aberration. This is because the electron beam interacts with the central region of the LG beam, where the influence of higher-order terms in the Taylor expansion of the ponderomotive potential is minimal. Consequently, any fifth-order contributions were below our detection sensitivity within the experimental conditions used. 


\subsubsection*{\label{sec:C}OFEM Spatial Intensity Distribution Reconstruction from the Measured Gradients}

Reconstructing a scalar field such as the intensity $I(x, y)$ from its spatial gradients is a classical inverse problem in imaging physics. Let $I(x, y)$ denote a twice-differentiable scalar field whose gradient is known:
\begin{equation}
    \nabla I(x, y) = \left( \frac{\partial I(x, y)}{\partial x}, \frac{\partial I(x, y)}{\partial y} \right).
\end{equation}

To recover the original intensity, one computes the divergence of this gradient field, which yields the Laplacian of $I(x, y)$:
\begin{equation}
    \nabla^2 I(x, y) = \nabla \cdot \left( \nabla I(x, y) \right).
\end{equation}

Applying the two-dimensional Fourier transform to both sides of the equation and using the identity that the Fourier transform of the Laplacian is equivalent to multiplication by the spatial frequency variables in the $x$ and $y$ directions, $-(k_x^2 + k_y^2)$, we obtain
\begin{equation}
    -(k_x^2 + k_y^2) \, \mathcal{F} \{ I(x, y) \} = \mathcal{F} \left\{ \nabla \cdot \left( \nabla I(x, y) \right) \right\}.
\end{equation}

Finally, taking the inverse Fourier transform provides the reconstructed intensity distribution in real space:
\begin{equation}
    I(x, y) = \mathcal{F}^{-1} \left\{ \frac{ -\mathcal{F} \left[ \nabla \cdot ( \nabla I(x, y) ) \right] }{ k_x^2 + k_y^2 } \right\}.
\end{equation}

To avoid division by zero at the origin of Fourier space, where $k_x = k_y = 0$, the corresponding component of the Fourier transform is set to zero manually. This removes the undefined global offset, which cannot be recovered from gradient data alone.

The U4DSTEM experiment was conducted using a relatively long objective lens working distance of 15.5 mm, with a 64 $\mu$m diameter aperture and the microscope operating at its highest current setting. Under these conditions, the electron beam radius at focus is limited by aberrations in the electron optics to about 20 nm, which defines the intrinsic spatial resolution. However, in the measurements presented in Fig. 3 in the main manuscript, the effective resolution was constrained by the scan step size of the electron focal spot, which was about 300 nm. The SLM was used to correct for deviations from the ideal LG intensity distribution. The long acquisition time required for the data in Fig.~\ref{fig:Donut Gradient}C ($30\,\mathrm{minutes}$) made it impractical for use in an active feedback loop. Nevertheless, since the electron beam interacts only with the central region of the LG beam (radius $<3\,\mu\mathrm{m}$), the profile in this area can be approximated as aberration-free.

\end{document}